\documentclass[
    ,final            
    ,numberedheadings 
  ]
  {aipproc}

\layoutstyle{6x9}

\def\lesssim{\mathrel{\hbox{\rlap{\hbox{\lower3pt\hbox{$\sim$}}}{\lower-1pt\hbox{$<$}}}}}
\def\gtrsim{\mathrel{\hbox{\rlap{\hbox{\lower3pt\hbox{$\sim$}}}{\lower-1pt\hbox{$>$}}}}}

\begin{document}
\title[Magnetic Acceleration and Collimation of GRB Jets]
{Magnetic Acceleration and Collimation of\\
Gamma-Ray Burst Jets}
\author{Arieh K\"onigl}{
  address={Department of Astronomy \& Astrophysics, University
  of Chicago, 5640 S. Ellis Ave.,\\
  Chicago, IL 60637, U.S.A.}
}


\begin{abstract}
Exact semianalytic solutions for GRB outflows were recently
derived using the equations of special-relativistic ideal MHD (see the
contribution by Vlahakis \& K\"onigl for a summary). This
contribution focuses on the implications of these
results to various modeling and observational issues in GRB sources, including
the baryon loading problem, polarization measurements of the prompt and reverse-shock
emission, and the possible existence of a two-component outflow.
\end{abstract}

\maketitle


\section{Magnetic Driving of GRB Outflows}

Gamma-ray burst (GRB) outflows are
likely powered by the extraction of rotational energy from a
newly formed stellar-mass black hole or neutron
star, or from a surrounding debris disk established in the
course of the central object's formation
\citep[e.g.,][]{M02}. Magnetic fields threading the central
object or disk provide the most plausible means of extracting
the inferred amount of energy on the timescale of the burst; they can
also guide, collimate, and accelerate the flow \citep[see][and
references therein]{VK01}. This picture is supported by a recent
measurement of a high ($80\pm20\%$) linear polarization in the
prompt $\gamma$-ray emission from GRB 021206 \citep{CB03}, which
can be plausibly interpreted in terms of a large-scale magnetic field advected
from the origin \citep[e.g.,][]{G03} and is consistent with
magnetic driving by an ordered field that threads the
source. Although purely hydrodynamic driving powered by neutrino emission
or magnetic field dissipation at the source can probably be
ruled out \citep[e.g.,][]{DPN02,DM02}, thermal effects may
nonetheless dominate the initial acceleration of magnetic jets
\citep[e.g.,][]{MLR93,VK01}.\footnote{It has also been argued
\citep[e.g.,][]{DS02} that electromagnetic energy dissipation could
contribute to the conversion of Poynting energy into kinetic energy
throughout the acceleration region of such flows.}

Motivated by the above considerations, Vlahakis \& K\"onigl
\citep[][hereafter VK03a and VK03b, respectively]{VK03a,VK03b}
constructed a general formalism for special-relativistic ideal MHD, allowing for
the presence of a baryonic component as well as of a ``hot''
electron-positron/radiation component that can dominate the
pressure. They showed how one can derive exact semianalytic solutions for
axisymmetric outflows under the assumption of radial self-similarity
and presented illustrative results for representative
GRB parameters. Vlahakis, Peng, \& K\"onigl \citep[][hereafter
VPK03]{VPK03} further generalized this scheme by obtaining solutions
for initially neutron-rich outflows, which they used to address the
baryon loading problem in GRB source models. A general description of
the formalism and of the derived solutions is given in
Vlahakis \& K\"onigl's contribution in these Proceedings. The
present contribution provides a brief overview of the main results and focuses
on their observational implications.
\section{Relativistic MHD Solutions}
The initial (subscript {i}) magnetic field amplitude can be
inferred from an estimate of the injected energy,
${\cal{E}}_{\rm i}$ = (Poynting flux) $\times$ (surface area) $\times$ (burst
duration). In a disk geometry [with initial cylindrical radius $\varpi_{\rm
i}$ and radial width $(\Delta \varpi )_{\rm i}$],
${\cal{E}}_{\rm i} \approx c E_{\rm i} B_{\phi,{\rm i}}
\varpi_{\rm i} (\Delta \varpi )_{\rm i} \Delta t$,
where the electric field is given by $E= B_{\rm p} V_\phi/c - B_\phi V_p/c$
(with the subscripts p and $\phi$ denoting the poloidal and
azimuthal components, respectively). For characteristic parameter
values [${\cal{E}}_{\rm i}\approx 10^{52}\ {\rm ergs}$,
$\varpi_{\rm i}\sim (\Delta \varpi )_{\rm i}\approx 10^6\ {\rm
cm}$, $\Delta t \approx 10\ {\rm s}$], one obtains $B_{\rm i} \sim
10^{14}-10^{15}\ {\rm G}$. This field is most plausibly generated by
differential-rotation amplification of a much weaker poloidal
field component that originally threads the source.

If $|B_{\rm p,i}/B_{\rm \phi,i}| > 1$, a {\em trans-Alfv\'enic} outflow
is produced, whereas if $|B_{\rm \phi,i}/B_{\rm p,i}| > 1$, the outflow is
{\em super-Alfv\'enic} from the start. The latter situation may correspond to
amplified toroidal flux loops that have been disconnected by
magnetic reconnection and escape from the disk surface in a nonsteady
fashion. Exact solutions for these two situations were derived in VK03a and VK03b,
respectively. It was demonstrated that, in either case, Poynting
flux-dominated jets can transform $\gtrsim 50\%$ of their
magnetic energy into baryon kinetic energy (with $E_{\rm K}
\sim 10^{51}\ {\rm ergs}$ and terminal Lorentz factors
$\gamma_{\infty} \sim 10^2-10^3$). If relativistic $e^+e^-$
pairs and radiation dominate the initial enthalpy, then a
thermal acceleration zone develops
at the base of the flow and remains dominant until the specific
enthalpy drops below $\sim c^2$, at which point magnetic
acceleration takes over. In contrast to the trans-Alfv\'enic solutions,
part of the enthalpy flux in the super-Alfv\'enic flows is
transformed into Poynting flux during the thermal acceleration
phase. Furthermore, the subsequent, magnetically dominated
acceleration in these flows can be significantly less rapid than
in the trans-Alfv\'enic case. 

The derived solutions have a free parameter, $F$, which controls
the distribution of the poloidal current
$I=c \varpi B_{\phi}/2$. For $F>1$ the flow is in the
current-carrying regime, with the poloidal current density being
antiparallel to the magnetic field. In this case the
current tends to zero as the symmetry axis is approached, so
such solutions should provide a good representation
of the conditions near the axis of a highly collimated
flow. Conversely, solutions with $F<1$ correspond to
the return-current regime (in which the poloidal current density
is parallel to the field) and are most suitable at larger cylindrical
distances. Although the detailed global current distribution
cannot be modeled using the self-similarity approach, one can
nevertheless generate ``hybrid'' flow configurations
that combine a current-carrying solution for low values of
$\varpi$ and a return-current solution for high values of
$\varpi$ (see Fig. 1 below for an example). Initially
Poynting-dominated flows that attain a
rough equipartition between the kinetic and Poynting energy fluxes at
large distances from the origin have $F$ close to 1. When $F>1$
the Lorentz force can collimate the flow to cylindrical asymptotics. For $F<1$ the
collimation is weaker and the flow only reaches conical
asymptotics; however, the acceleration is more efficient in this
case in that a larger fraction of the Poynting flux is converted into
kinetic energy.
\section{Implications to the Baryon Loading Problem}
As an illustration  of the unique properties of the relativistic
MHD solutions, consider the ramifications of a hydromagnetic jet
model to the baryon loading problem in GRB outflows. The
apparent difficulty stems from a comparison between the estimated
mass of protons in the jet, $M_{\rm proton}=3\times
10^{-6}(E_{\rm K}/10^{51}\ {\rm ergs})(\gamma_{\infty}/200)^{-1}\
M_\odot$, and the minimum mass of the debris disk from which the
jet is thought to originate, obtained under the assumption that
at most $\sim 10\%$ of the disk gravitational potential energy
could be converted into outflow kinetic energy. This comparison
implies that the outflow can comprise at most $\sim 10^{-4}$ of
the disk mass, whereas disk outflow models that utilize a large
fraction of the disk potential energy typically also entail
substantial mass loading. One approach to this issue has been to
postulate that the outflow emerges along magnetic field lines
that thread the black-hole event horizon and not the disk, but
then the converse problem --- how to avoid having too few
baryons --- must be addressed \citep[e.g.,][]{LE03}. A possible
resolution of the problem in the context of disk-fed jet models
was proposed in \citep{FPA00}, where it was noted that such outflows are
expected to be neutron-rich \citep[neutron/proton ratios as high as
${\rm n/p} \sim 20-30$; e.g.,][]{PWH03,B03b,VPK03}. Since
only the charged outflow component
couples to the electromagnetic field, the neutrons could
potentially decouple from the protons before the latter attain
their terminal Lorentz factor. In this picture, the inferred
value of $M_{\rm proton}$ may represent only a small fraction of the
total baryonic mass ejected from the disk, which would alleviate
the loading problem. However, it can be shown that, for purely
hydrodynamic outflows, the Lorentz factor $\gamma_{\rm d}$ at decoupling
is at least a few times $10^2$
\citep[e.g.,][]{DKK99,B03b,VPK03}. This implies that $\gamma_{\rm
d}/\gamma_{\infty} \sim 1$ and hence that the protons end
up with only a small fraction of the injected energy, which is
{\em not} a satisfactory resolution of the problem.

As demonstrated by VPK03, the incorporation of magnetic fields
makes it possible to attain $\gamma_{\rm d} \ll \gamma_{\infty}$
and thereby reclaim the promise of the Fuller et al. proposal. 
They wrote down the equations of motion for the neutron
component (which couples to the protons through a collisional
drag) and for the charged component (incorporating protons and
their neutralizing electrons as well as initially ``hot'' pairs
and radiation), and simplified them by considering a
well-coupled neutral/charged fluid for $\gamma \le \gamma_{\rm
d}$ and only the charged fluid component for $\gamma > \gamma_{\rm
d}$. The pre-decoupling region was described by a
super-Alfv\'enic outflow solution. As noted in \S~2, in
this case part of the enthalpy flux is converted into Poynting
flux during the initial thermal acceleration phase. This reduces
the acceleration rate, so at the point of decoupling
(when $V_{\rm proton}-V_{\rm neutron} \sim c$) the Lorentz
factor is still comparatively low. The energy deposited into the
Poynting flux is returned to the matter beyond the decoupling
point as kinetic energy, thereby enhancing the acceleration efficiency of
the proton component. The end result is a large
$\gamma_{\infty}/\gamma_{\rm d}$ ratio {\em and} comparable terminal
kinetic energies in the proton and neutron components, in clear
contradistinction to the purely hydrodynamic solutions.
\begin{figure}
\includegraphics[height=.5\textheight]{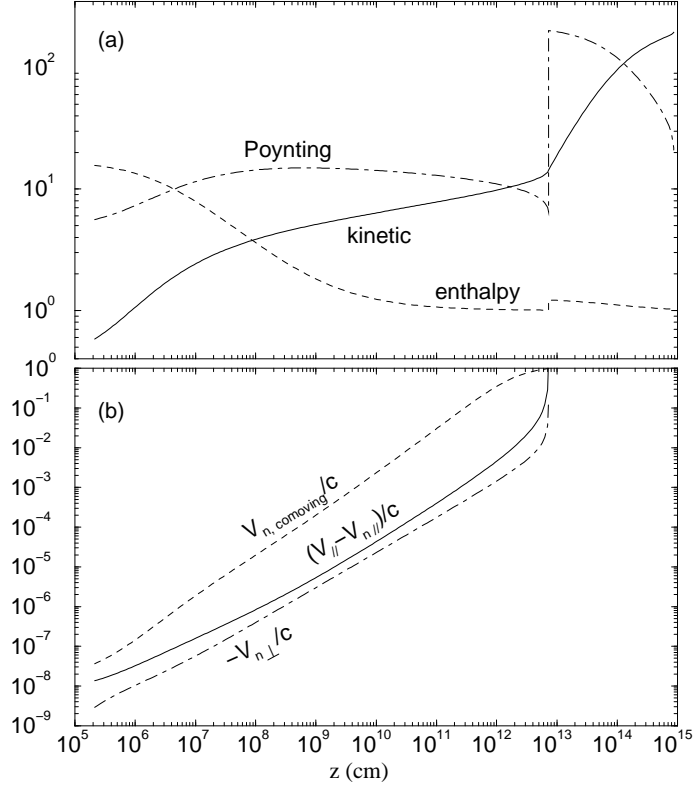}
\caption{Illustrative relativistic-MHD solution of a neutron-rich outflow. ($a$)
Components of the total energy flux, normalized by the mass flux
$\times\ c^2$, as functions of height along a fiducial magnetic field
line. The Poynting and enthalpy curves are discontinuous at the
decoupling point, reflecting the decrease in the mass flux of
field-coupled gas above that point. ($b$) Components of the
proton--neutron drift velocity.}
\end{figure}

An illustrative solution is shown in Fig. 1.\footnote{In this
example ${\rm n/p}=30$,
the pre-decoupling and post-decoupling regions correspond to the
current-carrying ($F=1.05$) and return-current ($F=0.1$)
regimes, respectively, and the flow collimates from an initial
opening half-angle of $55^{\circ}$ to $\theta_{\rm j} \approx 20^{\circ}$.}
The top panel shows the behavior of the various components of
the energy flux, corroborating the qualitative description given
above. The thermal acceleration effectively terminates at a
height $z\approx 10^9\ {\rm cm}$ above the disk, and the
neutrons decouple from the protons at $z_{\rm d} \approx
10^{13}\ {\rm cm}$, corresponding to $\gamma_{\rm d} \approx 15$.
By the time of decoupling the neutrons have acquired $\sim 2/3$ of the
injected energy, with the remainder residing predominantly in
the electromagnetic field. The latter portion is then
transferred with almost 100\% efficiency into proton kinetic
energy, so that, ultimately, the protons have $\gamma_{\infty} =
200$ and $E_{\rm K,proton} \approx 10^{51}\ {\rm ergs} \approx 0.5\,
E_{\rm K, neutron}$. The proton jet thus carries $\sim 1/3$ of the
injected energy but only $\sim 3\%$ of the injected mass.
The lower panel of Fig. 1 shows that, even though the decoupling
in this case is initiated by the growth of the n--p drift
velocity along the poloidal magnetic field,
there is also a transverse drift component (induced by the ongoing
magnetic collimation), which at the time of decoupling is
$V_{\rm neutron,\perp} \sim 0.1\, c$.\footnote{The exact value
of the angle between ${\mathbf{V}}_{\rm n}$ and
${\mathbf{V}}_{\rm p}$ at decoupling can
only be obtained by solving the equations of motion
without the ``strong coupling'' approximation adopted in the
solution shown in Fig. 1.}

\section{Additional Implications}
\noindent {\bf Polarization} --- Magnetic driving of GRB
outflows by large-scale, ordered magnetic fields would naturally lead to
a large linear polarization $P$ in the prompt $\gamma$-ray emission,
and a high value of $P$ is also predicted for the
emission from the {\em reverse shock} (the ``optical flash'' and
``radio flare'') \citep{GK03}. As shown in \citep{G03}, in this picture 
the prompt emission may be expected to exhibit $P \sim
43\%-61\%$ for typical values of the synchrotron-radiation
spectral index, consistent with the observations of GRB 021206 \citep{CB03}.
The ordered field is also expected to induce measurable circular
polarization \citep{MI03}.

\noindent {\bf Two-Component Outflow} --- The decoupled neutrons
in a neutron-rich outflow will undergo $\beta$ decay into protons at
a distance $\sim 4\times 10^{14}\, (\gamma_{\rm d}/15)\ {\rm
cm}$.  In contrast with the situation in purely hydrodynamic
outflow models \citep{PD02,B03a}, there may well be
{\em no} interaction between the two decoupled components
in the MHD case since their motions are not collinear (see Fig. 1$b$).
The latter scenario thus gives rise to a 2-component outflow: an
outer (wider) component (comprising the decoupled neutrons) that
carries most of the energy and may be responsible (after the
neutrons decay) for the bulk of the optical/radio afterglow, and an inner
(narrower) component (comprising the original protons) that
accounts for the prompt $\gamma$-rays and possibly also for much
of the X-ray afterglow. A 2-component outflow of this type
was inferred in GRB 030329 \citep{B03,S03}. A more detailed
investigation of this scenario is currently under way. If
$E_{\rm K,narrow} \lesssim E_{\rm K,wide}$ and $\theta_{\rm
j,narrow}/\theta_{\rm j,wide} \lesssim 1/3$, this picture would
make it possible to reconcile current inferences of the radiated
$\gamma$-ray energy \citep[e.g.,][]{BFK03} with internal-shock models.



\end{document}